\documentclass[runningheads]{llncs}

\usepackage{accv}

\usepackage{accvabbrv}

\usepackage{graphicx}
\usepackage{booktabs}

\usepackage{graphicx}
\usepackage{amsmath}
\usepackage{amssymb}

\usepackage{makecell}
\usepackage{subcaption}
\usepackage{caption}
\usepackage{subcaption}

\usepackage[accsupp]{axessibility}  

\usepackage{hyperref}

\usepackage{orcidlink}

\begin{document}

\title{Unsupervised Skull Segmentation via Contrastive
MR-to-CT Modality Translation} 

\titlerunning{Unsupervised Skull Segmentation via MR-to-CT CUT}

\author{Kamil Kwarciak\inst{1}\orcidlink{0000-0002-1392-4291} \and
Mateusz Daniol\inst{1}\orcidlink{0000-0003-2363-7912} \and
Daria Hemmerling\inst{1}\orcidlink{0000-0002-2193-7690} \and Marek Wodzinski\inst{1,2}\orcidlink{0000-0002-8076-6246}}

\authorrunning{K. Kwarciak et al.}

\institute{Department of Measurement and Electronics, AGH University of Krakow, Kraków, Poland \and Institute of Informatics, University of Applied Sciences Western Switzerland (HES-SO Valais), Sierre, Switzerland}

\maketitle

\begin{abstract}
The skull segmentation from CT scans can be seen as an already solved problem. However, in MR this task has a significantly greater complexity due to the presence of soft tissues rather than bones. Capturing the bone structures from MR images of the head, where the main visualization objective is the brain, is very demanding. The attempts that make use of skull stripping seem to not be well suited for this task and fail to work in many cases. On the other hand, supervised approaches require costly and time-consuming skull annotations. To overcome the difficulties we propose a fully unsupervised approach, where we do not perform the segmentation directly on MR images, but we rather perform a synthetic CT data generation via MR-to-CT translation and perform the segmentation there. We address many issues associated with unsupervised skull segmentation including the unpaired nature of MR and CT datasets (contrastive learning), low resolution and poor quality (super-resolution), and generalization capabilities. The research has a significant value for downstream tasks requiring skull segmentation from MR volumes such as craniectomy or surgery planning and can be seen as an important step towards the utilization of synthetic data in medical imaging.
 
\end{abstract}

\section{Introduction}
\label{sec:intro}

\begin{figure}[ht]
\begin{center}
    \centering
    \includegraphics[width=\textwidth]{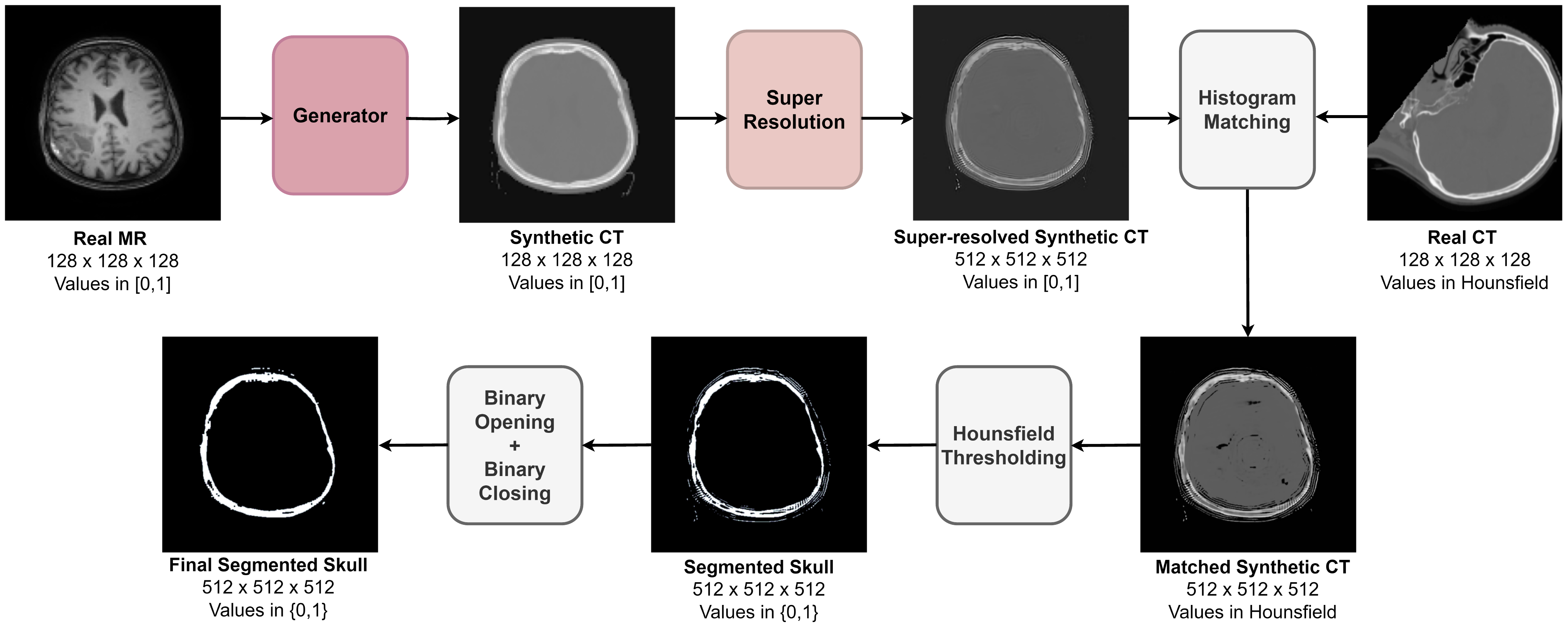}
    \caption{Inference with the use of the proposed solution. Firstly, the generator from the CUT framework is applied. This is followed by a super-resolution module. Next, we perform histogram matching using a different CT sample from the dataset, which does not need to be correlated with the input MR image. After this, Hounsfield thresholding is applied, followed by binary operations, to achieve the final skull segmentation results. The entire process leverages synthetic CT for skull segmentation via MR-to-CT modality translation.}
    \label{fig:inference}
\end{center}
\end{figure}

Skull segmentation is a common objective of many medical imaging studies, as it is usually the first step in processing pipelines and plays a crucial role in various diagnostic solutions. Among them, it is safe to mention: dental planning, patient positioning, or cranial implants design~\cite{wodzinski2021improving}. On one hand, this process in computed tomography (CT) may seem straightforward, as it commonly utilizes the thresholding in the Hounsfield scale~\cite{rulaningtyas2021ct}, while on the other hand, when it comes to magnetic resonance (MR) modality, this procedure gets more complicated. The reason behind this is that the main strength of MR lies in providing detailed images of soft tissues rather than bone structures. Hence, the problem becomes highly complex, but also indispensable, as many challenges require working on the skull (or in general bone) data, while only MR images are available or the acquisition of CT is undesired. As an example, one can mention the case of pediatric imaging where MR is favored over CT to avoid of ionizing radiation~\cite{bosch2023risk}. 

Among the techniques used for the separation of skull and brain structures, skull stripping~\cite{fisch2023deepbet,smith2002fast,hoopes2022synthstrip} seem to be standard solutions. Skull stripping is a technique that involves removing non-brain tissues, such as the skull and extracranial structures. This is typically achieved through a wide range of different methods such as classical morphological operations, or novel deep neural network architectures. Then, to obtain a skull segmentation mask, one, can perform a subtraction between the input MR image and output of skull stripping and apply binary masking on top of it. However, it should be kept in mind that skull stripping methods are not directly dedicated to skull segmentation, and they may produce some artifacts. Another approach that can be seen as a potential go-to solution is the utilization of foundation models, specifically Segment Anything Model~\cite{kirillov2023segment} and to be precise, its medical variant, MedSAM~\cite{ma2024segment}. Due to the robust MedSAM training routine which also involves training on a large-scale dataset that can handle diverse segmentation tasks, segmenting the skull from an MR image should also be possible. Finally, the skull segmentation models could be trained in a supervised manner, however, such an approach would require a significant number of annotated volumes from various medical centers and acquisition protocols.

The concept of modality translation has already emerged in the field of generative AI, as a very wide range of models and architectures have been developed. The general concept of image-to-image translation is to find a mapping between images from the source and target domains. The popular techniques include GAN-based approaches~\cite{isola2017image,zhu2017unpaired}, contrastive learning~\cite{park2020contrastive}, or diffusion models~\cite{sasaki2021unit}. Many of these solutions are applied in the medical image analysis to translate between modalities, where the most important one from the perspective of this work is MR-to-CT~\cite{wolterink2017deep,pan2023synthetic}. Existing solutions in this field yield satisfying outcomes. Nevertheless, their limitation lies in their inability to generalize effectively due to a fact of training on relatively homogeneous datasets. This causes a challenge for skull segmentation, especially in the context of clinical treatments, where data sources vary considerably. To take the matter further we find two major bottlenecks of contemporary strategies. Firstly, many approaches use solely 2-D data~\cite{wolterink2017deep,liu2021ct} rather than fully adopting a 3-D perspective, which is computationally cheaper, however limits inference performance in real-life scenarios with volumetric data, particularly when considering information between volumes. Secondly, various models are trained in the paired manner~\cite{pan2023synthetic} which means that there is a direct label and alignment between images from both domains in the dataset, and in most of the clinical cases, the data is available only in one modality. One more concept that should also be mentioned in terms of working on MR-to-CT translation is the resolution of these imaging techniques. Most commonly, MR data has a lower resolution than CT data, and high-resolution CT is desired in surgical procedures, such as craniectomy. Hence, in translation works it is important to further target obtained synthetic CT images and upsample them into higher-resolution, with the use of classic interpolation techniques or more robust, super-resolution networks.

\textbf{Contribution}: In this study, we aim to explore the emerging field of synthetic data generation for medical imaging, specifically focusing on the application of modality translation in cranioplasty to achieve unsupervised skull segmentation in MR images. We solve the previously mentioned dilemmas by designing a pipeline that enables the training on highly diverse datasets of MR and CT images in a fully volumetric and unpaired manner with the use of contrastive learning. We additionally provide a super-resolution module to further address opportunities in clinical use cases, and finally, we experiment with downstream tasks, such as the generation of synthetic CT skulls of children, and modality translation on defective skulls. We compare the proposed approach to state-of-the-art skull stripping methods with postprocessing and medical segmentation foundation model, MedSAM. Importantly, our model does not require any annotations during training. We would also like to emphasize the difference between skull segmentation and skull stripping. The primary goal of skull stripping is to separate the brain from non-brain elements, including the skull. If the final output needs to include the skull, additional postprocessing steps will be required. In contrast, the goal of skull segmentation is to specifically extract the cranial bone, making non-bone structures unimportant. The tasks are entirely different and face distinct challenges.

\section{Methods}
\subsection{Pipeline Overview}
\begin{figure}[ht]
\begin{center}
    \centering
    \includegraphics[width=\textwidth]{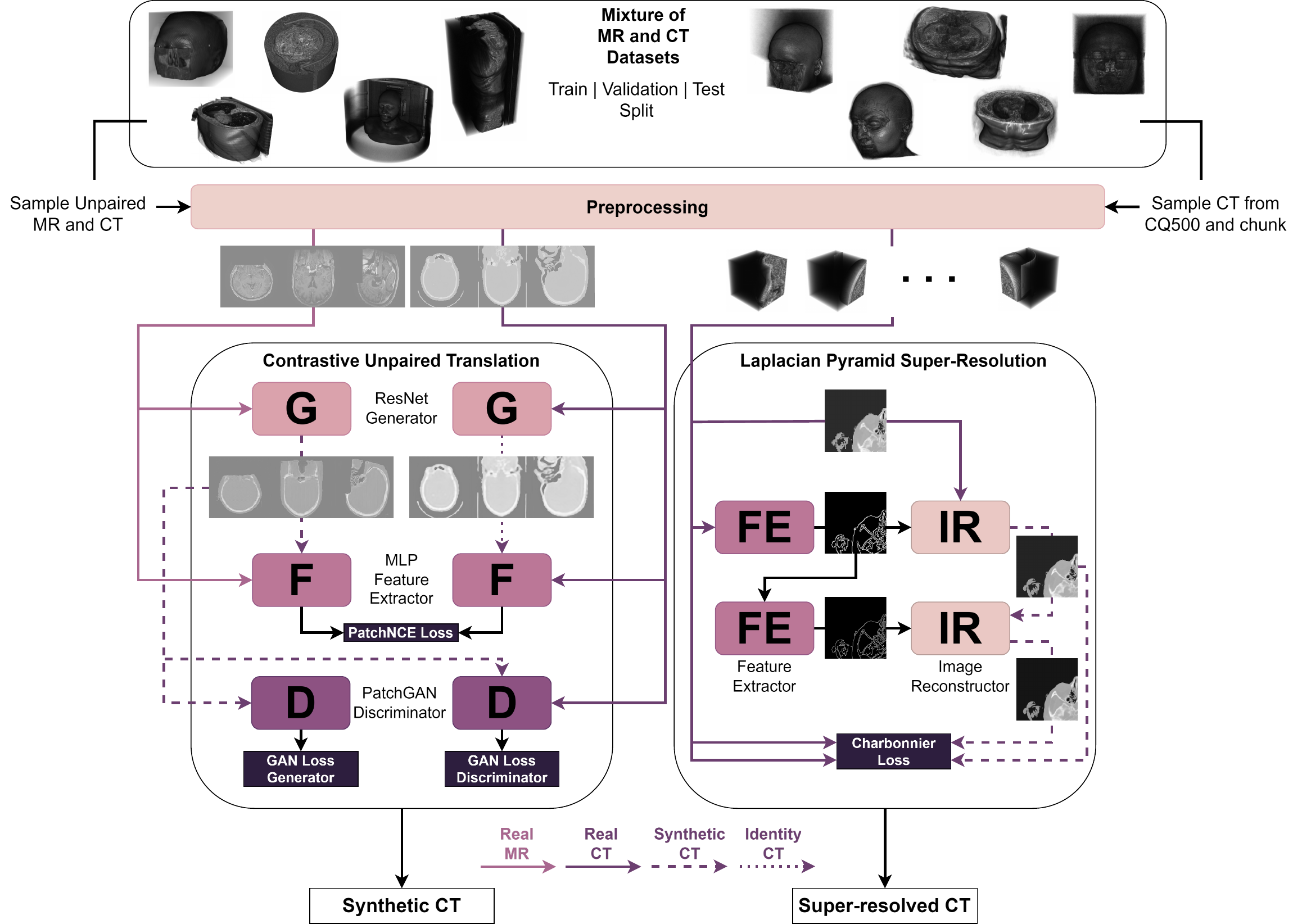}
    \caption{The overview of the pipeline for the Contrastive Unpaired Translation and Laplacian Pyramid Super-Resolution Network. Note that we work on 3-D tensors, the 2-D representations are used only for visualization simplicity.}
    \label{fig:pipeline}
\end{center}
\end{figure}

The two main components of our solution are the Contrastive Unpaired Translation (CUT)~\cite{park2020contrastive} module and the Laplacian Pyramid Super-Resolution Network (LapSRN)~\cite{lai2017deep} module, both utilized for 3-D volumetric data processing. They work separately and use different techniques for sampling the data from the whole space of gathered datasets. CUT is responsible for the MR-to-CT translation in an unpaired manner, it takes unpaired samples and tries to generate a synthetic CT image for the corresponding MR image without a vision of a direct mapping. LapSRN is trained directly and only on high-resolution CT images to create a pyramidal structure of increasing resolutions, with the usage of extreme data augmentation (which we discuss further) to enhance the generalization capabilities of the model. The full architecture is presented in Figure \ref{fig:pipeline} and the inference process is presented in Figure \ref{fig:inference}.

\subsection{Preprocessing}
For MR data we simply perform min-max normalization for every instance separately to obtain values in the range of [0,1] with preservation of the local dynamics. For CT data, we perform Hounsfield rescaling to remove the structures that have values below -500 $HU$ (air/background), as it enhances the training process, and has been found to be the most efficient for the whole dataset. After that, we also apply min-max scaling to values in the [0,1] range.

\subsection{Contrastive Unpaired Translation}
The CUT component of the pipeline is heavily inspired by the original work~\cite{park2020contrastive}, however, we add several updates to address the domain specificity. The CUT framework consists of three networks, generator $G$, discriminator $D$, and feature extractor $F$, where each of them is implemented to operate on 3-D data representations. $G$ in the context of this work is a ResNet-like encoder-decoder network, $D$ is a PatchGAN discriminator~\cite{isola2017image}, and $F$ is a shallow multilayer perceptron. In general, $G$ and $D$ follow a classic GAN minimax game, where $G$ aims to produce realistic CT data to fool a $D$ network, while the $D$ aims to distinguish between real and generated CT data:

\begin{align}
\label{eq:1}
   \mathcal{L}_{\text{GAN}}(D,G,\text{MR},\text{CT}) &= \mathbb{E}_{\mathbf{y}_{\text{real}}^{\text{CT}} \sim \text{CT}} \left[ \log D(\mathbf{y}_{\text{real}}^{\text{CT}}) \right] \\
   &\quad + \mathbb{E}_{\mathbf{x}_{\text{real}}^{\text{MR}} \sim \text{MR}} \left[ \log \left( 1 - D(G(\mathbf{x}_{\text{real}}^{\text{MR}})) \right) \right], \notag
\end{align}
where $G$ as input takes a real MR image $\mathbf{x}_{\text{real}}^{\text{MR}}$ and produces synthetic CT image $\mathbf{y}_{\text{syn}}^{\text{CT}}$ by learning a mapping $G: \mathbf{x}_{\text{real}}^{\text{MR}} \rightarrow \mathbf{y}_{\text{syn}}^{\text{CT}}$.
The second component of the CUT training objective is the use of noise contrastive estimation (NCE)~\cite{oord2018representation}. Specifically, it employs the InfoNCE loss to maximize mutual information by distinguishing the positive sample from a set of unrelated noise samples. However, an important aspect of the CUT framework, in comparison to the original InfoNCE work, is that the comparison is not performed between a positive sample and noise samples. Instead, it is done by comparing patches sampled from different spatial locations. 
The source MR image $\mathbf{x}_{\text{real}}^{\text{MR}}$, together with its corresponding synthetic CT image $\mathbf{y}_{\text{syn}}^{\text{CT}}$, are passed through the encoding component of the generator $G_{\text{enc}}$, followed by the feature extractor $F$, to generate latent feature vector representations in a shared embedding space. Next, a reference patch $\mathbf{z}_{\text{syn}}^{\text{CT}(\text{ref})}$ is extracted from the intermediate representation of $\mathbf{y}_{\text{syn}}^{\text{CT}}$ and compared to a positive patch $\mathbf{z}_{\text{real}}^{\text{MR}(+)}$ extracted from the intermediate representation of $\mathbf{x}_{\text{real}}^{\text{MR}}$ at the same spatial location, as well as to $N - 1$ negative samples $\mathbf{z}_{\text{real}}^{\text{MR}(-)}$ also taken from the intermediate representation of $\mathbf{x}_{\text{real}}^{\text{MR}}$, but taken from different spatial locations. This setup yields the following formulation of InfoNCE loss, with dot product used as a similarity measure:
\begin{align}
\label{eq:2}
    \ell_{\text{MR}} &= \mathcal{L}_{\text{InfoNCE} | \text{MR}_{\text{real}}, \text{CT}_{\text{syn}}} \big(\mathbf{z}_{\text{syn}}^{\text{CT}(\text{ref})}, \mathbf{z}_{\text{real}}^{\text{MR}(+)}, \mathbf{z}_{\text{real}}^{\text{MR}(-)} \big) \\ &= 
    - \log \frac{\exp \big( \mathbf{z}_{\text{syn}}^{\text{CT}(\text{ref})} \cdot \mathbf{z}_{\text{real}}^{\text{MR}(+)} \big)}{\exp \big( \mathbf{z}_{\text{syn}}^{\text{CT}(\text{ref})} \cdot \mathbf{z}_{\text{real}}^{\text{MR}(+)} \big) + \sum_{j=1}^{N-1} \exp \big( \mathbf{z}_{\text{syn}}^{\text{CT}(\text{ref})} \cdot \mathbf{z}_{\text{real}}^{\text{MR}(-)} \big)} \notag.
\end{align}

\begin{figure}[!t]
\begin{center}
    \centering
    \includegraphics[width=\textwidth]{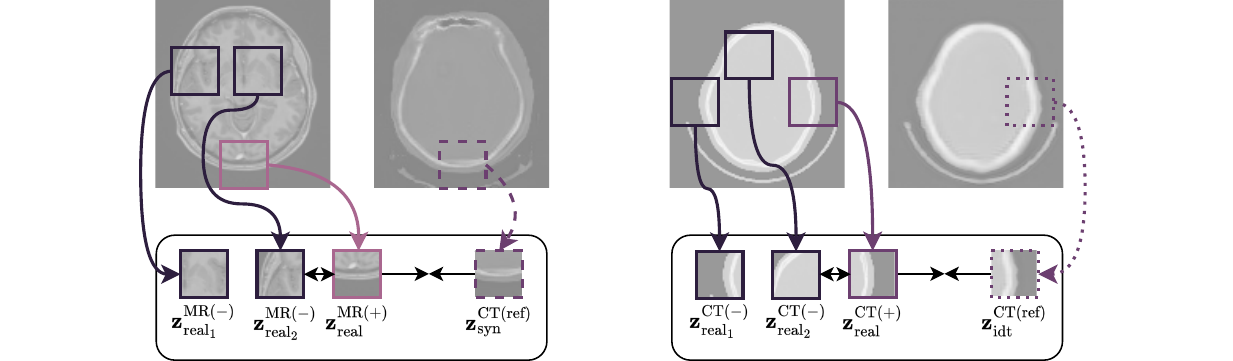}
    \caption{Sampling procedure for patchwise contrastive estimation of real MR $\leftrightarrow$ synthetic CT, and real CT $\leftrightarrow$ identity CT. We show a 2-D view for better visualization.}
    \label{fig:cut}
\end{center}
\end{figure}

The CUT methodology also incorporates the concept of identity mapping and utilizes InfoNCE in this setup to stabilize training, functioning similarly to an identity loss. These identity CT mappings are obtained via $ G: \mathbf{y}_{\text{real}}^{\text{CT}} \rightarrow \mathbf{y}_{\text{idt}}^{\text{CT}} \approx \mathbf{y}_{\text{real}}^{\text{CT}} $. The latent representation generation process is applied in a similar manner: the data is passed through the encoding part of the generator $ G_{\text{enc}}$ and then through the feature extractor $F$. This process results in the following set of patches: $\ \mathbf{z}_{\text{idt}}^{\text{CT}(\text{ref})}$, which is a reference patch from identity CT, and $\mathbf{z}_{\text{real}}^{\text{CT}(+)}$, $\mathbf{z}_{\text{real}}^{\text{CT}(-)}$ being positive and negative samples from real CT, this provides a second InfoNCE formulation as:
\begin{align}
\label{eq:3}
    \ell_{\text{CT}} &= \mathcal{L}_{\text{InfoNCE} | \text{CT}_{\text{real}}, \text{CT}_{\text{idt}}} \big(\mathbf{z}_{\text{idt}}^{\text{CT}(\text{ref})}, \mathbf{z}_{\text{real}}^{\text{CT}(+)}, \mathbf{z}_{\text{real}}^{\text{CT}(-)} \big) \\ &= 
    - \log \frac{\exp \big( \mathbf{z}_{\text{idt}}^{\text{CT}(\text{ref})} \cdot \mathbf{z}_{\text{real}}^{\text{CT}(+)} \big)}{\exp \big( \mathbf{z}_{\text{idt}}^{\text{CT}(\text{ref})} \cdot \mathbf{z}_{\text{real}}^{\text{CT}(+)} \big) + \sum_{j=1}^{N-1} \exp \big( \mathbf{z}_{\text{idt}}^{\text{CT}(\text{ref})} \cdot \mathbf{z}_{\text{real}}^{\text{CT}(-)} \big)} \notag.
\end{align}
The idea of comparison behind Equation \ref{eq:2} and \ref{eq:3} is presented in Figure \ref{fig:cut}. Moreover, the CUT methodology broadens the mentioned InfoNCE loss components to the intermediate outputs of $G_{\text{enc}}$ layers. This ensures that the features learned at multiple levels are aligned and collectively enhance the quality and consistency of the final generated output. If we denote the layers of the encoding part of $G$ ($G_{\text{enc}}$) as $L$, then at the $l$-th layer, we obtain a set of features for the real MR and real CT images as follows: $\{\mathbf{z}_{\text{real} (l)}^{\text{MR}} \}_{L} = \{F_l(G_{\text{enc}}^l(\mathbf{x}_{\text{real}}^{\text{MR}}))\}_L$ and $\{\mathbf{z}_{\text{real} (l)}^{\text{CT}} \}_{L} = \{F_l(G_{\text{enc}}^l(\mathbf{y}_{\text{real}}^{\text{CT}}))\}_L$. If we also address the synthetic and identity representations we obtain the following sets: $\{\mathbf{z}_{\text{syn} (l)}^{\text{CT}} \}_{L} = \{F_l(G_{\text{enc}}^l(G(\mathbf{x}_{\text{real}}^{\text{MR}})))\}_L$ and $\{\mathbf{z}_{\text{idt} (l)}^{\text{CT}} \}_{L} = \{F_l(G_{\text{enc}}^l(G(\mathbf{y}_{\text{real}}^{\text{CT}})))\}_L$. Now, as we take $S_l$ patches at layer $l$ we have a set of $S/s$ negative patches $(-)$ and $s$ positive patch $(+)$, and we are able to formulate PatchNCE loss in two setups, the first one as the main training objective of synthesizing CT images from MR inputs:

\begin{align}
\label{eq:4}
    \mathcal{L}&_{\text{PatchNCE}}(G,F,\text{MR}) \\ &= 
    \mathbb{E}_{\mathbf{x}_{\text{real}}^{\text{MR}} \sim \text{MR}} \sum_{l=1}^L \sum_{s=1}^{S_{l}} \ell_{\text{MR}} \big(\mathbf{z}_{\text{syn} (l)}^{\text{CT}(\text{ref})(s)}, \mathbf{z}_{\text{real} (l)}^{\text{MR}(+)(s)}, \mathbf{z}_{\text{real} (l)}^{\text{MR}(-)(S/s)} \big) \notag,
\end{align}
and the second being a stabilization term between real CT and its identity CT (real CT passed through the generator):

\begin{align}
\label{eq:5}
    \mathcal{L}&_{\text{PatchNCE}}(G,F,\text{CT}) \\ &= 
    \mathbb{E}_{\mathbf{y}_{\text{real}}^{\text{CT}} \sim \text{CT}} \sum_{l=1}^L \sum_{s=1}^{S_{l}} \ell_{\text{CT}} \big(\mathbf{z}_{\text{idt} (l)}^{\text{CT}(\text{ref})(s)}, \mathbf{z}_{\text{real} (l)}^{\text{CT}(+)(s)}, \mathbf{z}_{\text{real} (l)}^{\text{CT}(-)(S/s)} \big) \notag.
\end{align}
To conclude, for CUT in terms of MR-to-CT translation, PatchNCE loss is computed in two manners, firstly by comparing $\mathbf{x}_{\text{real}}^{\text{MR}}$ with $\mathbf{y}_{\text{syn}}^{\text{CT}}$ as a main training objective, and secondly by comparing $\mathbf{y}_{\text{real}}^{\text{CT}}$ with $\mathbf{y}_{\text{idt}}^{\text{CT}}$ for training stabilization and regularization. Finally, we can formulate the final objective of CUT training as:

\begin{align}
\label{eq:6}
\mathcal{L}_{\text{CUT}} &= \lambda_{\text{GAN}} \mathcal{L}_{\text{GAN}}(G,D,\text{MR},\text{CT}) \\ &+ \lambda_{\text{syn}} \mathcal{L}_{\text{PatchNCE}}(G,F,\text{MR}) + \lambda_{\text{idt}} \mathcal{L}_{\text{PatchNCE}}(G,F,\text{CT}) \notag
\end{align}

\subsection{Laplacian Pyramid Super-Resolution}
The second component of the designed pipeline is the modification of the original LapSRN \cite{lai2017deep} used for super-resolution of 3-D CT volumes. It uses two sub-networks: feature extractor and image reconstructor, which progressively generate images at higher resolutions from the low-resolution ones. The feature extractor is responsible for extracting the features at a coarse (lower) level and generating feature maps at a finer (higher) level. It enhances the representation of the input by capturing important details. The image reconstructor upsamples the lower-resolution input and then via element-wise summation with residuals obtained from the feature extractor creates a higher-resolution output with improved visual quality. By its nature, progressive reconstruction provides task-dependent flexibility and adjustability, as by bypassing the pyramid at certain levels we can obtain representations at the resolution required for a given task. Furthermore, to address the requirement of high generalization capabilities, our LapSRN uses not only a set of diverse data augmentations (flipping, affine transformations, motion artifacts, blurriness, contrast), but is also trained on overlapping chunks of input tensors (see Figure \ref{fig:pipeline}). This configuration enhances the model's generalization capabilities, enabling it to effectively super-resolve single chunks and capture inconsistencies in various skulls, such as those with defects. Additionally, it significantly reduces memory requirements during training, as the chunks can be processed individually or combined into small micro-batches within the mini-batch. LapSRN uses Charbonnier loss, where for $M$ chunks of the input tensor, and pyramid of $L$ levels, the Charbonnier loss is defined as:
\begin{equation}
\label{eq:7}
\mathcal{L}_{\text{Charbonnier}}(\mathbf{y},\mathbf{x},\mathbf{r}) = \frac{1}{M} \sum_{m=1}^M \sum_{s=1}^L \sqrt{(\mathbf{y}_s - \mathbf{x}_s - \mathbf{r}_s)^2 + \epsilon^2},
\end{equation}
where for layer $s$, chunk residual is denoted as $\mathbf{r}_s$, upsampled lower-resolution chunk as $\mathbf{x}_s$ and ground truth higher-resolution chunk $\mathbf{y}_s$ and $\epsilon$ (set to small value like 10$^{-3}$) controls similarity to $L_1$ loss while staying differentiable.

\subsection{Postprocessing}
Synthetic CT and high-resolution CT obtained from pipeline components are in the [0, 1] range and it is desired to have them in the Hounsfield scale, to perform threshold-based skull segmentation. This can be easily achieved by performing a histogram matching with one of the real CTs from the train set for CUT, and low-resolution CT for LapSRN. Furthermore, it was experimentally found that following Hounsfield-based thresholding with binary opening and binary closing is beneficial for artifacts removal. It is also important to note, that training both modules, CUT and LapSRN on the original Hounsfield scale can yield training instabilities due to the wide range of values and potential outliers, which can lead to gradient issues and slow convergence.

\section{Experiments}
\subsection{Datasets}
To address the generalization possibilities we create a huge dataset of 1,521 MR and 879 CT 3-D images from publicly available datasets \cite{keane2021brain,rogers2023real,peelle2022increased,ottesen2020differences,fialkowski2022identifying,greene2018behavioral,burleigh2023fear,banfi2021reading,gaesser2019role,jo2020brain,thornton2019people,etzel2022dual,racey2023open,thummerer2023synthrad2023,podobnik2023han,radl2022avt,chilamkurthy2018deep,kavur2021chaos} (See Supplementary materials for more information). Importantly, we extract a portion of the dataset from the SynthRAD 2023 Challenge~\cite{thummerer2023synthrad2023} for testing purposes, as it is the only dataset that includes paired MR and CT samples, hence it can be used for evaluation with metrics like Dice coefficient. To address the requirements of generalization, we construct the dataset of skull segmentation by including also other anatomical structures, such as the pelvis, aorta, or kidneys. This setup enables a more robust capturing of style rather than context, as commonly different institutions use various imaging devices, making single-institution models hard to generalize. To address this, we combined datasets with images of different anatomical structures to enhance cross-institution generalizability via different anatomical structures and different acquisition settings. We train CUT with the use of all datasets, and for LapSRN we only use a dataset of high-resolution CT skulls, namely CQ500~\cite{chilamkurthy2018deep}.

\subsection{Networks}
We implement the proposed networks with the use of PyTorch library~\cite{paszke2019pytorch}. CUT and LapSRN subnetworks use 3-D convolutional layers. What's important, for the CUT model we use instance normalization~\cite{ulyanov2016instance} as it prevents instance-specific mean and covariance shifts, hence it is highly beneficial for modality translation tasks. CUT's generator is a 9-layer ResNet-based network, with residual connections between downsampling and upsampling blocks, it uses instance normalization and ReLU nonlinearity besides the decision layer for which we use hyperbolic tangent. The discriminator is a 3-layer convolutional network with Leaky ReLU and instance normalization (like PatchGAN discriminator~\cite{isola2017image}). CUT's feature extractor is a simple 2-layer multilayer perceptron with ReLU which operates on 64 patches of features extracted from the generator's flow. Hence, regarding Equation \ref{eq:4} and Equation \ref{eq:5}, we operate on a fixed amount of patches equal to 64, and a fixed amount of $G_{\text{enc}}$ layers equal to 9. LapSRN's image reconstructor is a 2-layer convolution/deconvolution network and the feature extractor is an 8-layer convolution/deconvolution network with Leaky ReLU nonlinearities. LapSRN's convolutional layers use 3$\times$3$\times$3 kernels with 64 filters with He initialization~\cite{he2015delving}, deconvolutions use the kernel of 4$\times$4$\times$4 (upsampling by a factor of 2) and weights are initialized from a trilinear filter, as suggested in the original implementation~\cite{lai2017deep}.

\subsection{Setup}
We resize the data to 128 $\times$ 128 $\times$ 128 for training both the CUT and LapSRN models. We train CUT using ADAM optimizer~\cite{kingma2014adam} with $\beta_1$ = 0.5 and $\beta_2$ = 0.999. We set the initial learning rate to 2 $\cdot$ 10$^{-4}$ and decrease it linearly every 50 epochs if the loss doesn't decrease. We set $\lambda_{\text{GAN}}$, $\lambda_{\text{syn}}$ and $\lambda_{\text{idt}}$ all equal to 1 and train CUT with the real batch size of 1 (batch size of 8 distributed across 8 GPUs). LapSRN is trained using SGD with momentum term set to 0.9 and weight decay of 10$^{-4}$. The learning rate is initialized to 10$^{-5}$ and decreases linearly every 5 epochs if the loss doesn't decrease. The chunk size for splitting the inputs is set to 8 with 8 voxels overlapping (128$^3$ $\rightarrow$ 8 $\times$ 64(+8)$^3$), the batch size is 1 and we accumulate gradients every 16 batches. Models were trained until convergence, with the use of 8 NVIDIA A100 40GB GPUs.

\section{Results}

\begin{table}[!t]
\caption{Comparison of methods with DSC and SDSC, evaluated on the subset of SynthRAD 2023 training dataset~\cite{thummerer2023synthrad2023}, extracted for a test set in this work.}
\label{tab:DSC_eval}
\centering
\begin{tabular}{lcc}
\multicolumn{3}{c}{} \\ \hline
Method          & DSC $\uparrow$    & SDSC $\uparrow$     \\ \hline
SynthStrip~\cite{hoopes2022synthstrip} + subtraction      & 0.281     & 0.580   \\
Bet~\cite{smith2002fast} + subtraction            & 0.273     & 0.563   \\
DeepBet~\cite{fisch2023deepbet} + subtraction        & 0.298     & 0.606   \\
MedSAM~\cite{ma2024segment}          & –        & –    \\
Ours            & \textbf{0.512}     & \textbf{0.770}
\end{tabular}
\end{table}

\begin{figure}[!b]
	\centering
    \begin{subfigure}{\columnwidth}
		\centering
		\includegraphics[width=\textwidth]{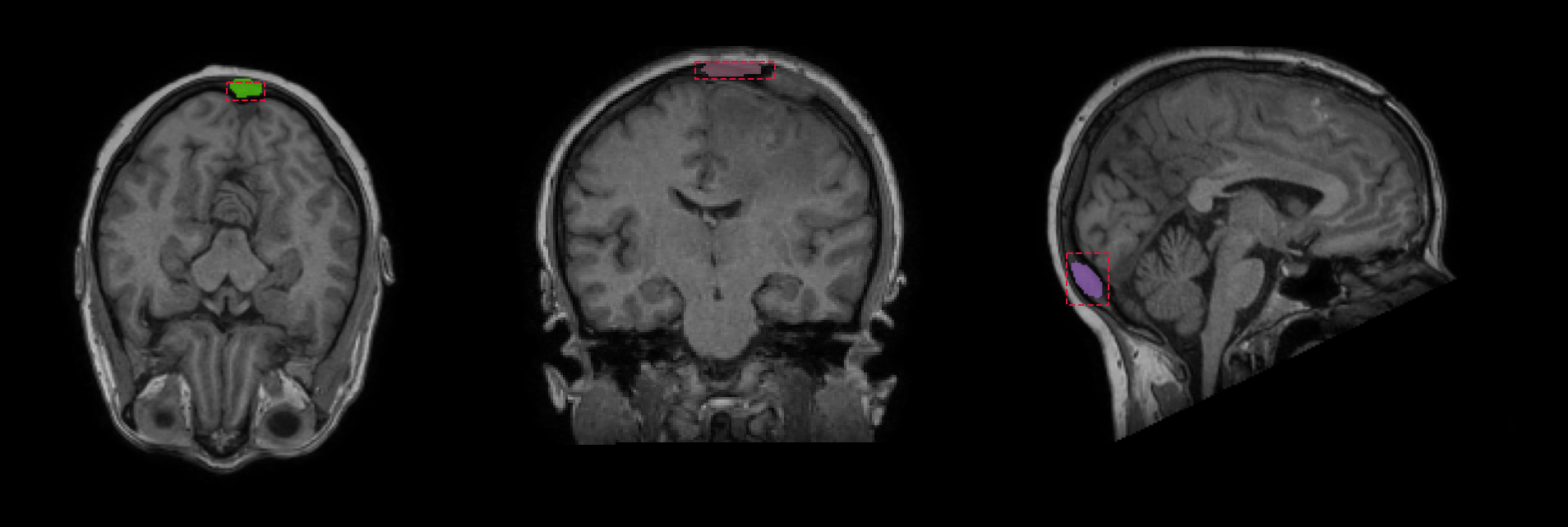}
		\subcaption{Bounding box approach}\label{bb}
	\end{subfigure}
	
    \begin{subfigure}{\columnwidth}
		\centering
        \includegraphics[width=\textwidth]{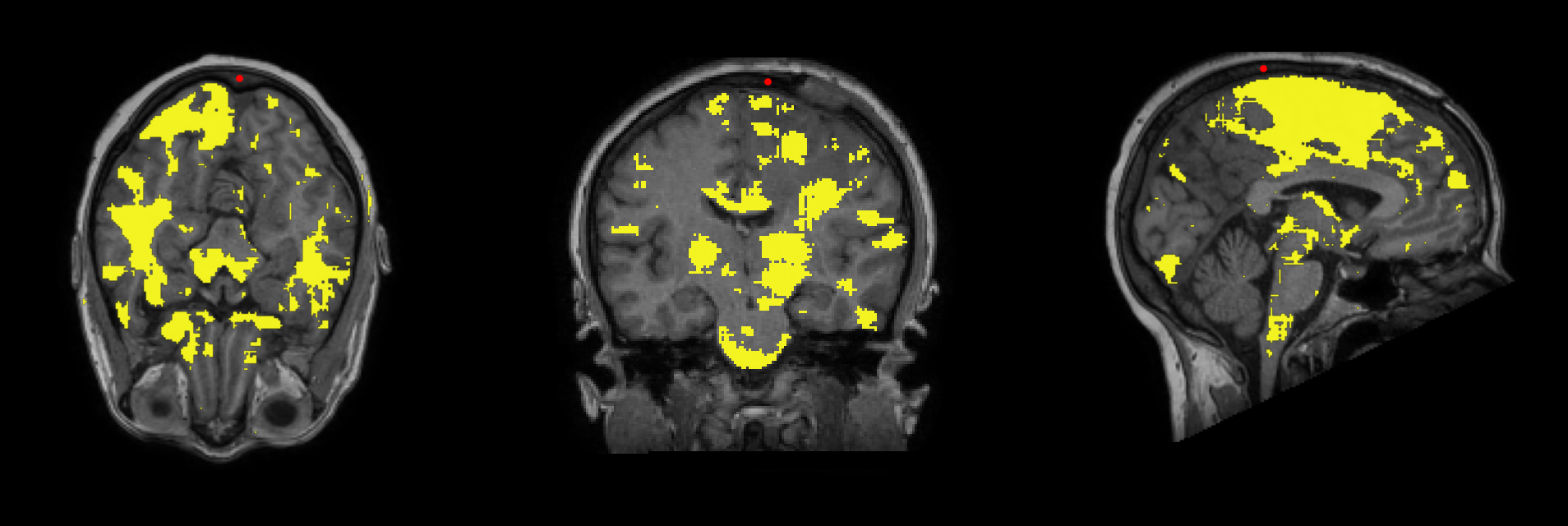}
		\subcaption{Point prompt approach}\label{pp}
	\end{subfigure}
\caption{Results of skull segmentation from MR images with the use of MedSAM: (\protect\subref{bb}) bounding box and (\protect\subref{pp}) point prompt approaches. Bounding box approach failure stems from a fact that segmented skull structure is not propagated through the whole image, and only small parts are captured. For point prompt, the model is unable to identify the skull and propagates segmentation into brain.}
\end{figure}

\begin{figure}[!t]
\begin{center}
    \includegraphics[width=\textwidth]{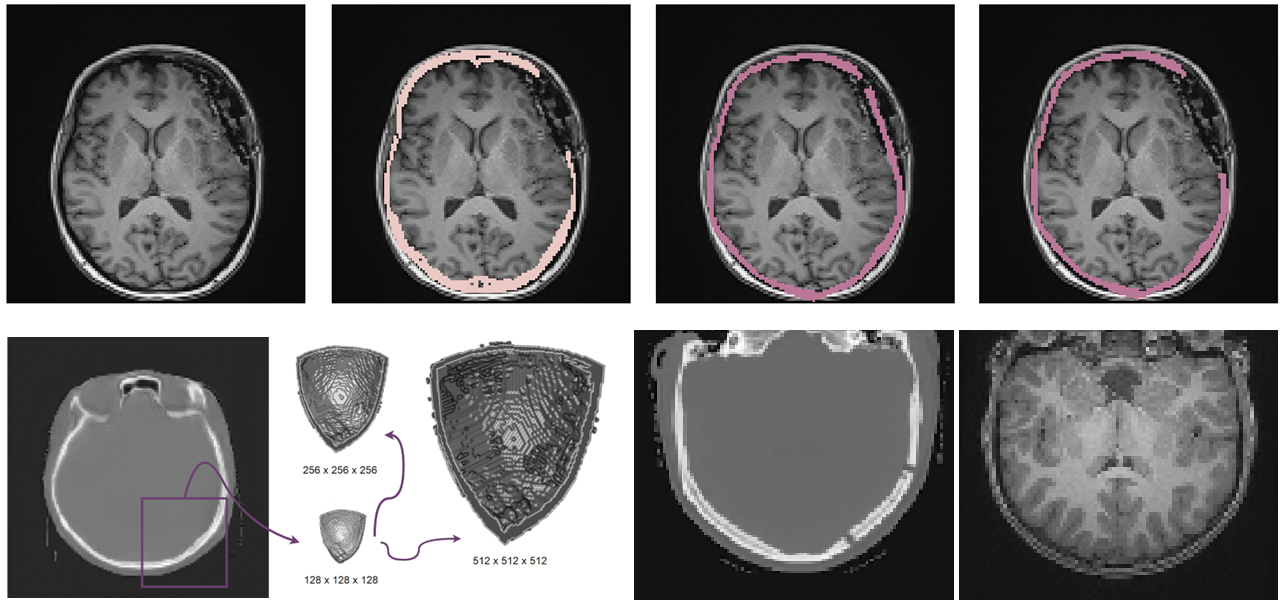}
    \caption{Top: Results of translation and segmentation on defected skulls (from left to right: input MR, matched CT mask, synthetic CT mask, synthetic CT mask with removed implant area). Bottom left: Super-resolution of synthetic skull. Bottom right: MR-to-CT translation on child's skull.}
    \label{fig:evals}
\end{center}
\end{figure}

We evaluate the proposed method on the subset of samples extracted from the training dataset from SynthRAD 2023 Challenge~\cite{thummerer2023synthrad2023} as it consists of paired MR and CT images. We present the Dice coefficient (DSC) and surface Dice coefficient (SDSC) results, calculated from segmentation masks obtained using Hounsfield scale thresholding in Table \ref{tab:DSC_eval}. Importantly, it should be noted that, as previously mentioned, skull stripping differs from skull segmentation. The results of skull stripping methods were followed by postprocessing, which involved manual thresholding for skull subtraction. Furthermore, we demonstrate that the medical imaging foundation model, MedSAM~\cite{ma2024segment}, fails to perform the complex task of skull segmentation from MR images. The failed results are presented in Figure \ref{bb} for the bounding box approach and Figure \ref{pp} for the experimental point prompt approach. We investigate the quality of generated synthetic CT via a small set of ablation studies which are presented in Figure \ref{fig:evals}. They showcase the following settings: $(i)$ the performance of LapSRN using a test sample from the CUT module, $(ii)$ generalization capabilities to the downstream task of translation on the MR image of the child's skull, and $(iii)$ the translation of the MR image of a defected (and reconstructed) skull. Finally, to demonstrate the generalization potential of the proposed solution from a quantitative perspective, we investigate additionally the MR subset of the HaN-Seg dataset~\cite{podobnik2023han} which was not present in the original training dataset. As the availability of paired MR and CT datasets is highly limited, besides SynthRAD 2023 dataset~\cite{thummerer2023synthrad2023}, we find HaN-Seg to be the most suitable option for this evaluation. Importantly, additional preprocessing in the form of image registration between CT and MR samples in this dataset was required to enable the quantitative evaluation. For comparison, we decided to train a state-of-the-art segmentation network, SwinUNETR~\cite{hatamizadeh2021swin}, on the SynthRAD 2023 dataset~\cite{thummerer2023synthrad2023}, where the input was an MR image and the target was a skull mask derived from Hounsfield thresholding of its paired CT image. We also trained the model as a standard paired MR-to-CT translation and applied the same Hounsfield thresholding on the resulting synthetic CTs for consistency with the pipeline from Figure \ref{fig:inference}. Table \ref{tab:Han} shows that the proposed solution presents superior results in terms of the generalization capabilities in comparison to both SwinUNETR setups.

\begin{table}[!b]
\centering
\caption{Evaluation of the generalization capabilities on the HaN-Seg dataset~\cite{podobnik2023han} in terms of DSC and SDSC.}
\label{tab:Han}
\begin{tabular}{lccc}
\hline
Method              & DSC $\uparrow$    & SDSC $\uparrow$       \\ \hline
SwinUNETR (MR-Mask) & 0.084  & 0.135        \\
SwinUNETR (MR-CT)   & 0.113  & 0.149        \\
Ours       & \textbf{0.310} & \textbf{0.396} \\
\end{tabular}
\end{table}

\section{Discussion}
Quantitative analysis of the CUT method shows that it can provide better results in terms of DSC and SDSC than the combination of skull stripping with subtraction and masking as shown in Table \ref{tab:DSC_eval}. Importantly, we find out that MedSAM~\cite{ma2024segment} struggles with the skull segmentation in MR images and fails to propagate the segmentation mask in both bounding-box and point-prompt (experimental) approaches. This can be potentially solved via task-specific fine-tuning (as the authors mention this MedSAM capability in their work). Good results of our method suggest that it is a promising direction for designing a fully unsupervised framework of skull segmentation from MR images, that can be used for downstream tasks (such as defect segmentation), and with the use of super-resolution LapSRN module, the requirement of high-resolution CT images for tasks like craniectomy or surgery planning is also met. What's more, both CUT and LapSRN are relatively simple in terms of architectural design and also their learning objectives, hence they can be adapted to other translation tasks, or fine-tuned for other anatomical structures. With CUT we were able to obtain good generalization capabilities due to the very high diversity of the used dataset. Importantly, we also note several limitations of our methodology and leave it as a potential direction for further work. First of all, produced results sometimes include some blurriness; this can be attributed to using convolutional networks with a relatively small number of filters and a shallow network depth. 
This issue stems from the challenges of designing large networks for high-resolution volumetric data. Even the latest GPUs struggle with VRAM capacities when trying to fit large tensors and extensive 3-D networks. Secondly, while DSC results are better than other methods, they still require improvement to meet the demands of potential real-world medical applications. Finally, the primary motivation of this work was to achieve generalization capabilities for the segmentation task in an unsupervised and unpaired manner. Therefore, we do not compare our results with other methods used in MR-to-CT translation, as these are not primarily focused on segmentation. Nevertheless, we acknowledge that some existing translation methods may be more effective than CUT, and other super-resolution modules could outperform LapSRN. We consider this a potential area for further improvement, research, and investigation.

In conclusion, we presented a novel generative AI-based methodology for synthetic CT generation, specifically designed for skull segmentation from MR images. Our approach operates on 3-D data, is trained in an unsupervised manner, and demonstrates strong generalization capabilities while also super-resolving into higher resolutions. We believe this opens up new research opportunities in this field, and we plan to further enhance the proposed solution.


\section*{Acknowledgements}
The project was funded by The National Centre for Research and Development, Poland under Lider Grant
no: LIDER13/0038/2022 (DeepImplant). We gratefully
acknowledge Polish HPC infrastructure PLGrid support
within computational grants no. PLG/2023/016239 and
PLG/2024/017079.

%
%
\bibliographystyle{splncs04}
\bibliography{main}

\clearpage
\pagenumbering{gobble}
\appendix 

\section*{Supplementary Materials - Datasets Details} 

\begin{table}[!b]
\caption*{\textbf{Table:} Information about datasets used in training of CUT and LapSRN modules.}
\label{tab:Data}
\begin{tabular}{|l|l|l|l|}
\hline
\textbf{Dataset Reference} & \textbf{Modality} & \textbf{Instances} & \textbf{License} \\ \hline
\makecell[l]{Keane \textit{et al.}~\cite{keane2021brain}, Dataset DOI: \\doi:10.18112/openneuro.ds003404.v1.0.1}   & MR       & 80                  & CC0      \\ \hline
\makecell[l]{Rogers \textit{et al.}~\cite{rogers2023real}, Dataset DOI: \\doi:10.18112/openneuro.ds004285.v1.0.0}            & MR       & 156                 & CC0      \\ \hline
\makecell[l]{Peelle \textit{et al.}~\cite{peelle2022increased}, Dataset DOI: \\doi:10.18112/openneuro.ds003717.v1.0.1}                & MR         & 120                    & CC0        \\ \hline
\makecell[l]{Ottesen \textit{et al.}~\cite{ottesen2020differences}, Dataset DOI: \\doi:10.18112/openneuro.ds003606.v1.0.0}                & MR         & 29                    & CC0     \\ \hline
\makecell[l]{Fialkowski \textit{et al.}~\cite{fialkowski2022identifying}, Dataset DOI: \\doi:10.18112/openneuro.ds003831.v1.0.0}                & MR         & 73                    & CC0        \\ \hline
\makecell[l]{Greene \textit{et al.}~\cite{greene2018behavioral}, Dataset DOI: \\doi.org/10.1016/j.neuroimage.2018.01.023}                & MR         & 32                    & PDDL        \\ \hline
\makecell[l]{Burleigh \textit{et al.}~\cite{burleigh2023fear}, Dataset DOI: \\doi:10.18112/openneuro.ds004393.v1.0.4}                & MR         & 30                    & CC0        \\ \hline
\makecell[l]{Banfi \textit{et al.}~\cite{banfi2021reading}, Dataset DOI: \\doi:10.18112/openneuro.ds003126.v1.3.1}                & MR         & 58                    & CC0        \\ \hline
\makecell[l]{Gaesser \textit{et al.}~\cite{gaesser2019role}, Dataset DOI: \\doi:10.18112/openneuro.ds001439.v1.2.0}                & MR         & 18                    & CC BY-SA 4.0        \\ \hline
\makecell[l]{Jo \textit{et al.}~\cite{jo2020brain}, Dataset DOI: \\doi:10.18112/openneuro.ds003103.v1.0.1}                & MR         & 30                    & CC0        \\ \hline
\makecell[l]{Thornton \textit{et al.}~\cite{thornton2019people}, Dataset DOI: \\doi:10.18112/openneuro.ds004217.v1.0.0}                & MR         & 53                    & CC0        \\ \hline
\makecell[l]{Etzel \textit{et al.}~\cite{etzel2022dual}, Dataset DOI: \\doi:10.18112/openneuro.ds003465.v1.0.6}                & MR         & 108                    & CC0        \\ \hline
\makecell[l]{Racey \textit{et al.}~\cite{racey2023open}, Dataset DOI: \\doi:10.18112/openneuro.ds004466.v1.0.1}                & MR         & 254                    & CC0        \\ \hline
\makecell[l]{Thummerer \textit{et al.}~\cite{thummerer2023synthrad2023}, Dataset DOI: \\doi.org/10.5281/zenodo.7260705}                & MR \& CT         & 360 \& 360                    & CC BY-NC 4.0        \\ \hline
\makecell[l]{Podobnik \textit{et al.}~\cite{podobnik2023han}, Dataset DOI: \\doi.org/10.1002/mp.16197}                & CT         & 42                    & CC BY-NC-ND 4.0        \\ \hline
\makecell[l]{Radl \textit{et al.}~\cite{radl2022avt}, Dataset DOI: \\doi.org/10.1016/j.dib.2022.107801}                & CT         & 38                    & CC BY 4.0        \\ \hline
\makecell[l]{Chilamkurthy \textit{et al.}~\cite{chilamkurthy2018deep}, Dataset DOI: \\doi.org/10.1016/S0140-6736(18)31645-3}                & CT         & 399                    & CC BY-NC-SA 4.0        \\ \hline
\makecell[l]{Kavur \textit{et al.}~\cite{kavur2021chaos}, Dataset DOI: \\doi.org/10.5281/zenodo.3362844}                & MR \& CT         & 120 \& 40                    & CC BY-NC-SA 4.0        \\ \hline
\end{tabular}
\end{table}

\end{document}